\documentclass[prl,twocolumn,amsmath,superscriptaddress,amssymb]{revtex4}

\usepackage{graphicx}
\usepackage{dcolumn}
\usepackage{bm}
\usepackage{amssymb}
\usepackage{amsmath}
\usepackage{wasysym}
\usepackage{color}
\usepackage{times}

\usepackage{placeins}
\usepackage{epstopdf}

\newcommand{\rv} {{\mathbf r}}

\newcommand{\bra}[1]{\left\langle\,#1\,\right|}
\newcommand{\ket}[1]{\left|\,#1\,\right\rangle}

\newcommand{\mme}[3]{\langle\,#1\,|#2|\,#3\,\rangle}

\newcommand{\yu} {{\chi^{\uparrow}  }}
\newcommand{\yd} {{\chi^{\downarrow}}}

\newcommand{\xu} {{\chi^{\uparrow}  }}
\newcommand{\xd} {{\chi^{\downarrow}}}

\usepackage[normalem]{ulem}

\makeatletter
\renewcommand*\env@matrix[1][c]{\hskip -\arraycolsep
  \let\@ifnextchar\new@ifnextchar
  \array{*\c@MaxMatrixCols #1}}
\makeatother
\begin{document}

\title{Strong Linear Dichroism in Spin-Polarized Photoemission from Spin-Orbit-Coupled Surface States}

\author{H. Bentmann}\affiliation{Experimentelle Physik VII and R\"ontgen Research Center for Complex Materials (RCCM), Universit\"at W\"urzburg, Am Hubland, D-97074 W\"urzburg, Germany}
\author{H. Maa\ss}\affiliation{Experimentelle Physik VII and R\"ontgen Research Center for Complex Materials (RCCM), Universit\"at W\"urzburg, Am Hubland, D-97074 W\"urzburg, Germany}
\author{E. E. Krasovskii}\affiliation{Departamento de F\'{i}sica de Materiales, 
Facultad de Ciencias Qu\'{i}imicas, Universidad del Pais Vasco/Euskal Herriko Unibertsitatea, 
Apdo. 1072, San Sebasti\'{a}n/Donostia, 20080 Basque Country, Spain}
\affiliation{Donostia International Physics Center (DIPC), Paseo Manuel de Lardizabal 4, 
San Sebasti\'{a}n/Donostia, 20018 Basque Country, Spain}
\affiliation{IKERBASQUE, Basque Foundation for Science, 48013 Bilbao, Spain}
\author{T. R. F. Peixoto}\affiliation{Experimentelle Physik VII and R\"ontgen Research Center for Complex Materials (RCCM), Universit\"at W\"urzburg, Am Hubland, D-97074 W\"urzburg, Germany}
\author{C. Seibel}\affiliation{Experimentelle Physik VII and R\"ontgen Research Center for Complex Materials (RCCM), Universit\"at W\"urzburg, Am Hubland, D-97074 W\"urzburg, Germany}
\author{M.~Leandersson}\affiliation{MAX IV Laboratory, Lund University, P. O. Box 118, 221 00 Lund, Sweden}
\author{T. Balasubramanian}\affiliation{MAX IV Laboratory, Lund University, P. O. Box 118, 221 00 Lund, Sweden}
\author{F. Reinert}\affiliation{Experimentelle Physik VII and R\"ontgen Research Center for Complex Materials (RCCM), Universit\"at W\"urzburg, Am Hubland, D-97074 W\"urzburg, Germany}
\date{\today}

\begin{abstract}
A comprehensive understanding of spin-polarized photoemission is
crucial for accessing the electronic structure of spin-orbit coupled materials. Yet, the impact of the final state in the photoemission process on the photoelectron spin has been difficult to assess in these systems. We present 
experiments for the spin-orbit split states in a Bi-Ag surface alloy showing that 
the alteration of the final state with energy may cause a complete reversal of the photoelectron
spin polarization. We explain the effect on the basis of {\it ab initio} one-step photoemission theory 
and describe how it originates from linear dichroism in the angular distribution of photoelectrons. Our 
analysis shows that the modulated photoelectron spin polarization reflects the intrinsic spin density 
of the surface state being sampled differently depending on the final state, and it indicates linear 
dichroism as a natural probe of spin-orbit coupling at surfaces.
\end{abstract}
\maketitle

The creation and manipulation of spin-polarized electronic states in
crystalline solids, low-dimensional systems, and heterostructures through 
strong spin-orbit interaction is a central topic in contemporary condensed matter physics
\cite{molenkamp:07,Hasan:10.11,heinz:14,Manchon:15,Bibes:16}. Among
the most vivid examples are the surface states of topological
insulators in which the coupling of the spin and momentum degrees of
freedom gives rise to helical spin textures in momentum space
\cite{Hsieh:09}. Moreover, spin-orbit split band structures, in
general, attract attention in a broad range of materials,
including Weyl semimetals \cite{Xu:15}, with unconventional
spin-polarized states in the bulk and at the surface,
strongly-correlated topological Kondo insulators
\cite{Coleman:10,Min:14}, as well as two-dimensional systems, such as
metallic oxide interfaces \cite{Bibes:16} and transition-metal
dichalcogenide layers \cite{heinz:14}. Thus, given the tremendous interest in spin-orbit coupled materials,
it is of critical importance to probe their electronic structure with spin sensitivity and to reliably verify the anticipated spin dependences, see, {\it e.g.}, Refs.~\cite{Xu:11,jozwiak:13,Donath:13,riley:14}.

The most versatile tool to spectroscopically address the
momentum-dependent spin polarization of electronic band structures in
condensed matter physics has been spin- and angle-resolved photoemission spectroscopy
(spin-ARPES). In recent years it has been successfully applied to a variety of materials
\cite{Dil:09,heinzmann:12,Okuda:13}. At the same time, the efficiency
of state-of-the-art photoelectron spin detectors has been improved
tremendously \cite{Okuda:13,jozwiak:13,Tusche:15}, making it now
possible to measure the photoelectron spin polarization over wide
regions of momentum space with varying energy and polarization of the
exciting light. Despite these encouraging developments fundamental
issues remain debated, namely to which degree, under which conditions,
and in which way the measured photoelectron spin polarization actually reflects the intrinsic spin properties of spin-orbit split
states \cite{jozwiak:13,Wissing:14,zhu:14,sanchez-barriga:14,Wortelen:15,krasovskii:15,Seibel:16,Miyamoto:16,maass:16,Noguchi:17}. 

Within the one-step theory of photoemission the spin-dependent photocurrent is
determined by the photoemission matrix element which involves the initial state of the 
transition -- the object of interest -- and the final state of the outgoing photoelectron. 
In the vacuum ultraviolet photon-energy regime, commonly used in
spin-ARPES experiments, the properties of the final state may quickly vary 
with the excitation energy and deviate considerably
from the naive assumption of a free-electron state~\cite{Krasovskii:07}. This is well-known to induce pronounced
modulations of the spin-integrated photoemission intensity, as observed, {\it e.g.}, 
for surface states on metals \cite{Mulazzi:09,ortega:11} and on topological insulators
\cite{Scholz:13}. Yet, a comprehensive understanding of how the final state affects the spin-resolved photocurrent of 
spin-orbit split states remains elusive. 

\begin{figure*}
\includegraphics[width=5.5in]{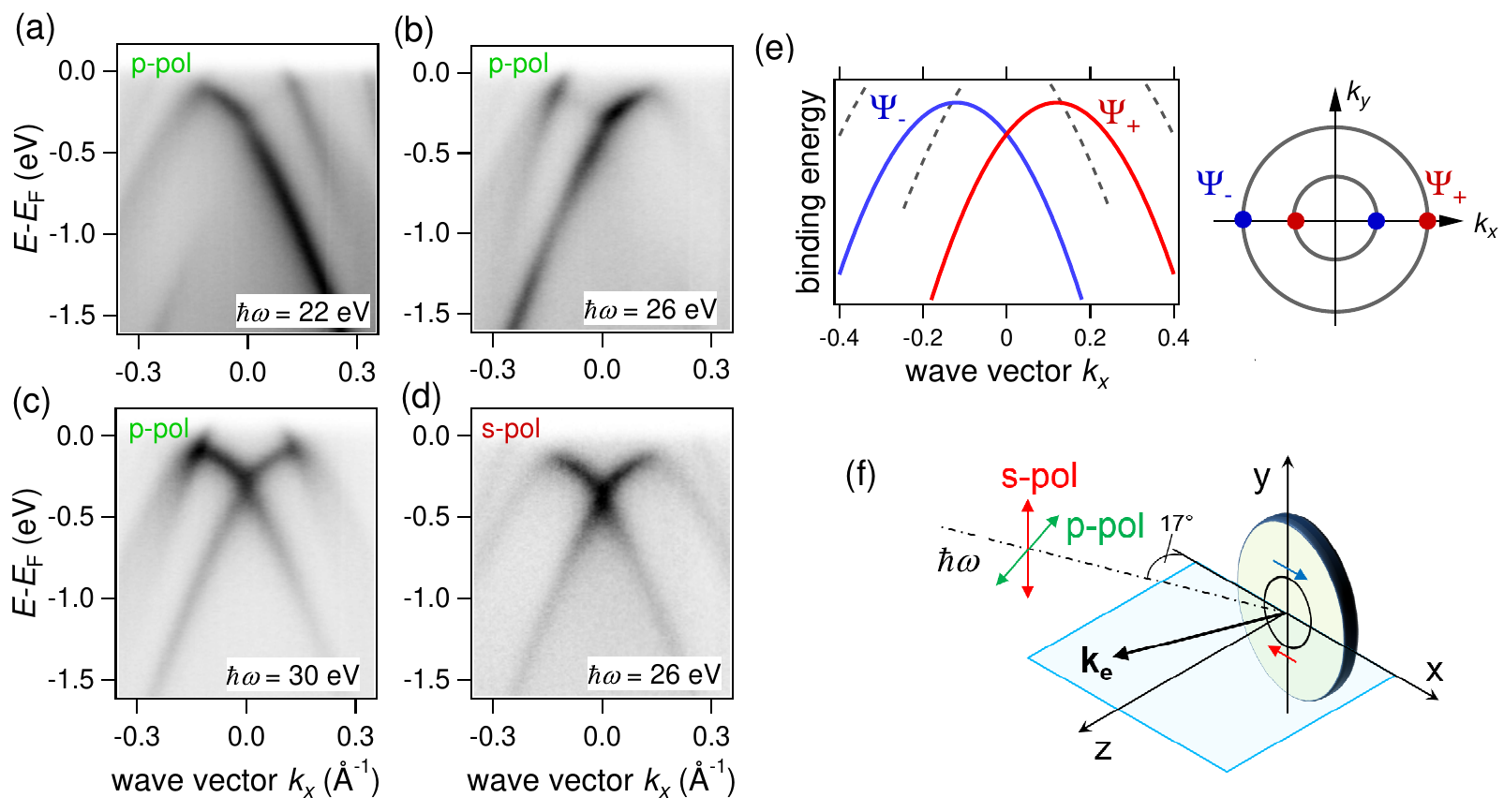} %
\caption{\label{fig1}Angle-resolved photoemission data of the spin-orbit split surface states in BiAg$_2$/Ag(111) taken along the $\bar{\Gamma}\bar{M}$ direction ($k_x$) using $p$-polarized in (a)-(c) and $s$-polarized light in (d). (e) Schematic of the spin-orbit split branches $\Psi_{\rm +}$ and $\Psi_{\rm -}$. (f) Sketch of the experimental geometry.}
\end{figure*}

Here, we show in a joint experimental and theoretical study that variation of the final-state wave function with energy can induce a complete reversal of the photoelectron spin. Based on general symmetry considerations and dipole selection rules we relate this surprising reversal in the photoelectron spin polarization to simultaneous modulations in the angle-resolved photoemission intensity. Both effects originate from a strong linear dichroism in the angular distribution of photoelectrons and are confirmed by {\it ab initio} one-step photoemission theory. The results are obtained for the surface alloy BiAg$_2$/Ag(111), a model system for spin-orbit effects \cite{ast:07,Bentmann:12.05,Bode:13,Mugarza:15} and spin-dependent photoemission \cite{Meier:11.4,meier:09,Chiang:12,Wissing:14,Noguchi:17}. Its electronic structure features a nearly parabolic, Bi 6$p$-derived band with negative effective mass. Due to large Rashba-type spin-orbit coupling the band is split into two branches $\Psi_{\rm +}$ and $\Psi_{\rm -}$ with opposite spin polarization [Fig.~\ref{fig1}(e)].          

The experiments were performed at room-temperature and in ultrahigh vacuum ($p<2\cdot10^{-10}$~mbar). The surface alloy \cite{ast:07,Meier:11.4} was grown as described elsewhere \cite{Moreschini:09}. Spin-ARPES experiments were performed at beamline I3 of the Max-lab storage ring (Lund, Sweden). The experimental geometry is depicted in Fig.~\ref{fig1}(c). We used a Scienta R4000 photoelectron analyzer with a Mott-detector operated at 25~kV. The energy resolution of the ARPES and spin-ARPES experiments were approximately 15~meV and 50~meV, respectively. The angular resolution for spin-ARPES experiments was 3$^{\circ}$. The Sherman function of the Th Mott-target was $S_{\rm eff}=0.17$.

We first consider spin-integrated ARPES data along the $\bar{\Gamma}\bar{M}$ direction ($k_x$) obtained with $p$-polarized light [Fig.~\ref{fig1}(a)-(c)]. The intensities of both branches $\Psi_{\rm +}$ and $\Psi_{\rm -}$ strongly modulate with photon energy [see also Fig.~\ref{fig2}(a) and Fig.~S1 in the supplement], in agreement with earlier work \cite{meier:09}. The dependence of the intensity on photon energy differs for $\Psi_{\rm +}$ and $\Psi_{\rm -}$. This leads to pronounced $\pm k_x$ asymmetries that can be classified as linear dichroism in the angular distribution of photoelectrons \cite{Henk:04}. They are related to the light electric field vector $\mathbf{E}=(\mathcal{E}_x,0,\mathcal{E}_z)$, which breaks the $\pm k_x$ mirror symmetry of the experimental setup. By contrast, symmetric intensity distributions are observed when using $s$-polarized light, as seen in Fig.~\ref{fig1}(d) and in Fig.~S2 of the supplement.

The ARPES data along $k_x$ are largely independent of the azimuthal crystal orientation [Fig.~S4 of the supplement]. We may, therefore, assume that the system, besides the mirror symmetry $x\to-x$, also has the mirror symmetry $y\to-y$. This is reasonable because the surface states are strongly localized in the alloy layer, which has both mirror reflections. For a non-degenerate state with ${\mathbf k}_\parallel$ in the mirror plane $xz$ the wave functions $\Psi_{\rm +}$ and $\Psi_{\rm -}$ along $k_x$, are connected by time-reversal symmetry and can be written as $\Psi_{\rm +}=g\yu + u\yd$ and $\Psi_{\rm -}=u^*\yu - g^*\yd$, with $\chi^{\sigma}$ quantized along $y$. The spinor component $g(\rv)$ is even under the reflection $y\to-y$, and $u(\rv)$ is odd \cite{Henk:03}. Hence, the net spin density $|g(\rv)|^{2} - |u(\rv)|^{2}$ varies with position $\rv$ as a consequence of spin-orbit coupling. We then introduce the mirrored functions $\tilde g(x,y,z)=g(-x,y,z)$ and $\tilde u(x,y,z)=u(-x,y,z)$ and write $\Psi_{\rm -}={\tilde u}\yu + {\tilde g}\yd$.

\begin{figure*}
\includegraphics[width=6.5in]{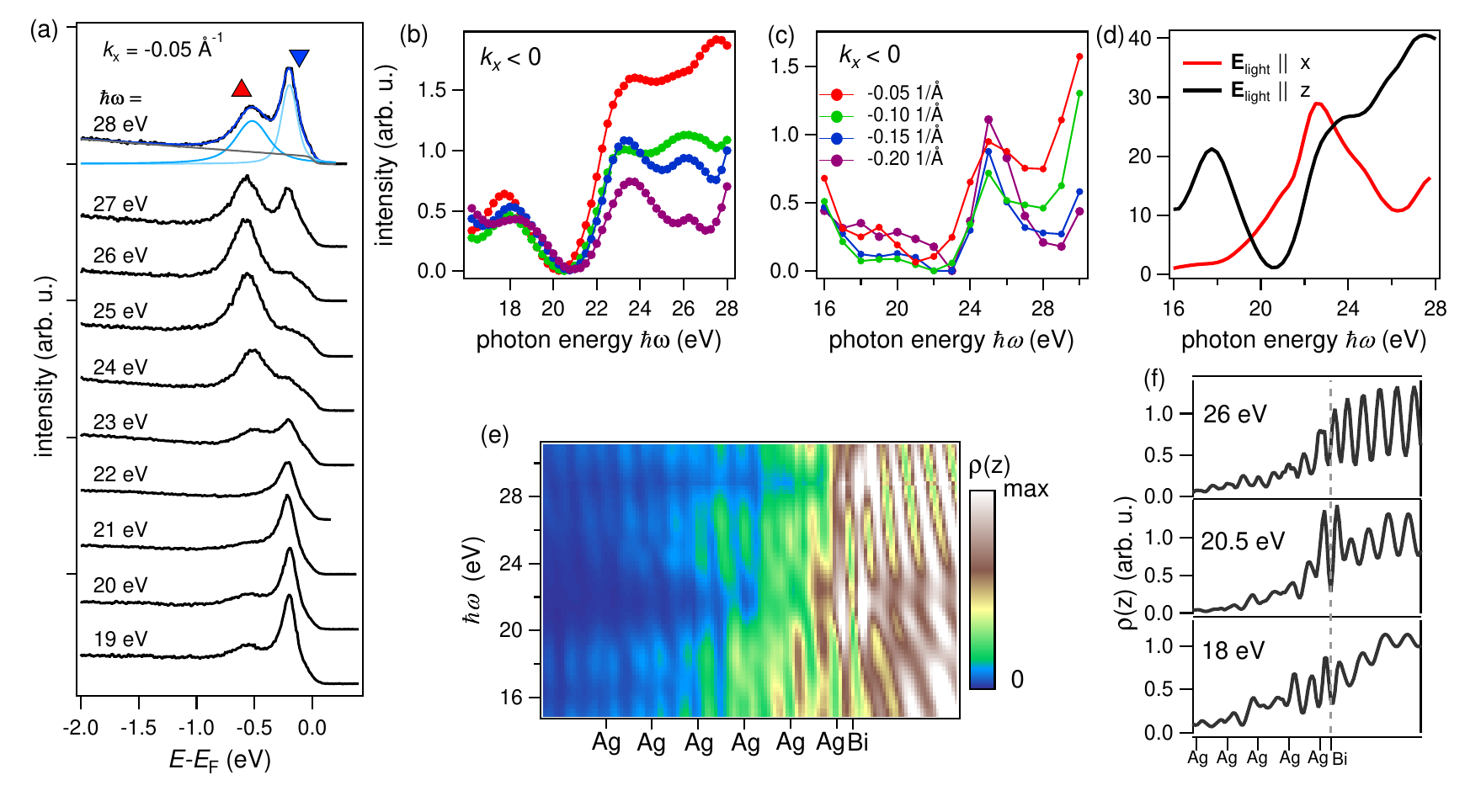} %
\caption{\label{fig2} Modulation of the photoemission intensity of the spin-orbit split branches $\Psi_{\rm +}$ and $\Psi_{\rm -}$ in BiAg$_2$/Ag(111) with photon energy (cf. Fig.~\ref{fig1}). Panel (a) shows energy distribution curves (EDC) at negative wave vectors $k_x$. The spectra were normalized to the background intensity at binding energies below 1.5 eV. The two peaks in each EDC are assigned to the branches $\Psi_{\rm +}$ and $\Psi_{\rm -}$. Calculated and measured intensities for $\Psi_{\rm +}$ depending on photon energy are 
shown in (b) and (c). (d) Calculated intensities $I_z$ and $I_x$ for the light electric field along $z$ and along $x$, respectively, for the branch $\Psi_{\rm +}$ at negative $k_x$. (e) Probability density $\rho(z)$ of the photoelectron final state $\ket{\Phi}$ as a function of the surface-normal coordinate $z$ and $\hbar\omega$. (f) Line profiles of $\rho(z)$ in (e) at selected photon energies.}
\end{figure*}  

Within the dipole approximation for the photoexcitation operator $\hat O$ rigorous 
parity selection rules hold: because the final state -- the time-reversed LEED 
state $\ket{\Phi}$ -- is necessarily even, for $p$-polarized light incident in the 
emission plane it is $\bra{\Phi}\hat O\ket{u}=0$, and for $s$-polarized light it is
$\bra{\Phi}\hat O\ket{g}=0$. Hence, for $p$-polarization the photoemission intensities $I_{\rm +}$ and $I_{\rm -}$ 
for $\Psi_{\rm +}$ and $\Psi_{\rm -}$ can be written as
\begin{align} 
I_{\rm +}&=|\mme{\Phi}{\mathcal{E}_zp_z}{g} + \mme{\Phi}{\mathcal{E}_xp_x}{g}|^2 {\rm~and}\\
I_{\rm -}&=|\mme{\tilde \Phi}{\mathcal{E}_zp_z}{\tilde g} + \mme{\tilde\Phi}{\mathcal{E}_xp_x}{\tilde g}|^2,
\end{align} 
where we have neglected the spin-orbit interaction in the final state. Because
$T_z=\mme{\Phi}{\mathcal{E}_zp_z}{g}=\mme{\tilde \Phi}{\mathcal{E}_zp_z}{\tilde g}$ and
$T_x=\mme{\Phi}{\mathcal{E}_xp_x}{g}=-\mme{\tilde \Phi}{\mathcal{E}_xp_x}{\tilde g}$ we 
obtain the intensity asymmetry: $I_{\rm +}=|T_z+T_x|^2$ and $I_{\rm -}=|T_z-T_x|^2$.

A full suppression of $I_{\rm +}$ and $I_{\rm -}$ is expected for $I_z \approx I_x$, with $I_{x,z}=|T_{x,z}|^2$, and for phase differences between $T_z$ and $T_x$ of $\Delta\phi =\pi$ and $\Delta\phi =0$, respectively. These conditions are met for $\hbar\omega =$~22~eV ($I_{\rm +}$~=~0) and $\hbar\omega =$~26~eV ($I_{\rm -}$~=~0) [Fig.~\ref{fig1}(a)-(b)]. A simultaneous zero-crossing of $T_z$ and $T_x$ can be excluded as, in this case, both $I_{\rm +}$ and $I_{\rm -}$ would be suppressed. Given the grazing angle of light incidence ($\mathcal{E}_z / \mathcal{E}_x \approx 3$) one would naturally assume $I_z > I_x$, leading to rather symmetric intensity distributions. This is indeed the case for most parts of the studied photon-energy range of $\hbar\omega =8...40$~eV [Fig.~S1 in the supplement]. However, near $\hbar\omega =$~22~eV and $\hbar\omega =$~26~eV we find $I_z\approx I_x$, and in between one may expect $I_x > I_z$.      

In order to go beyond these qualitative considerations we present one-step photoemission calculations of the 
$\hbar\omega$-dependent intensity. 
The final state $\ket{\Phi}$ is the time-reversed low energy electron 
diffraction state~\cite{Adawi64} calculated for the scattering of electrons 
on a slab composed of the surface alloy and five layers of Ag(111) substrate. 
The potential of the slab was obtained within the local density approximation 
with the full-potential linear augmented plane wave method (LAPW)~\cite{KSS99}. 
For the initial states the relativistic effects were included within a two-component 
formalism~\cite{KOH77}, and the final states were obtained in the scalar relativistic 
approximation with the inverse LAPW method~\cite{Krasovskii99}. A detailed description 
of the procedure can be found in Ref.~\cite{Borghetti2012}. The inelastic scattering of 
photoelectrons is included by adding a spatially constant imaginary part 
$V_{\rm i}=1$~eV to the crystal potential.

\begin{figure}
\includegraphics[width=3.3in]{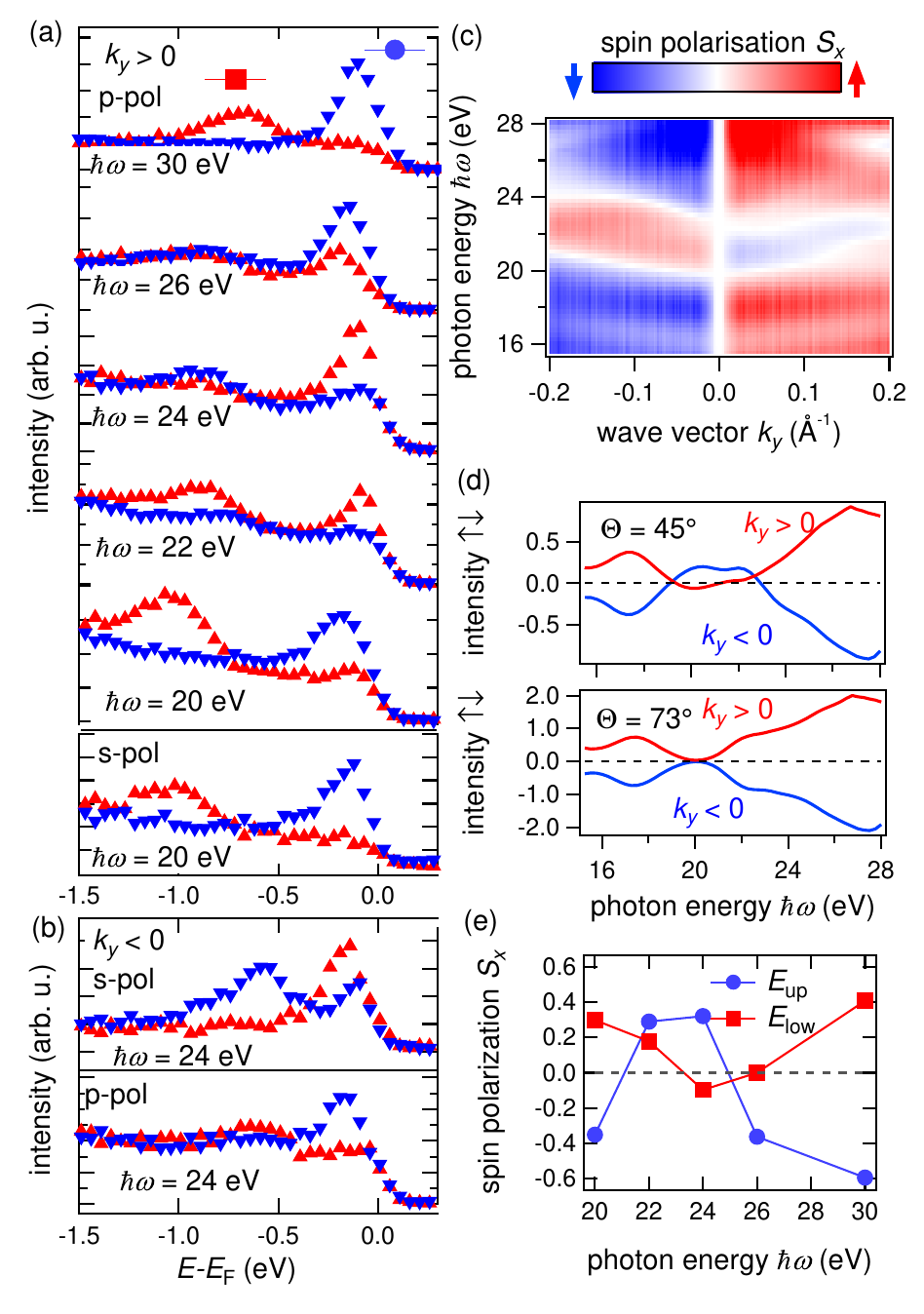} %
\caption{\label{fig3} Modulation of the photoelectron spin polarization in BiAg$_2$/Ag(111) with excitation energy 
$\hbar\omega$. (a)-(b) Spin-resolved EDC at positive (a) and negative (b) wave vectors along $k_y$ [cf. Fig.~\ref{fig1}(e)-(f)].
The spin-quantization axis of the measurement is aligned with the $x$-axis. (c) One-step photoemission 
calculation of the photoelectron spin polarization for the branch at lower energy along $k_y$ assuming $p$-polarized 
light and an angle of incidence $\theta =$~45$^\circ$. (d) Calculated net-spin photocurrents of the branch at lower energy at $k_y = \pm$0.1{\AA}$^{-1}$ for $\theta =$~45$^\circ$ and $\theta =$~73$^\circ$ (e) Measured photoelectron spin polarization for the spin-orbit split branches depending on $\hbar\omega$, as extracted 
from the data in (a).}
\end{figure}

Figure~\ref{fig2}(d) shows the calculated intensities $I_z$ and $I_x$ of the branch $\Psi_{\rm +}$ for light polarized linearly along $z$ and along $x$, respectively. We find markedly different $\hbar\omega$ dependences for $I_z$ and $I_x$ and, in particular, a pronounced minimum of $I_z$ near $\hbar\omega =$~21~eV as well as a maximum of $I_x$ near $\hbar\omega =$~23~eV. As expected from the experimental data, we find a range of photon energies where $I_x > I_z$. 

In Fig.~\ref{fig2}(b)-(c) we compare the calculated and measured photoemission intensities of the branch $\Psi_{\rm +}$. The strong intensity variation observed at these energies is remarkably well reproduced by the {\it ab initio} theory. It is related to a Bragg gap at approximately 22 eV in the unoccupied band structure of the Ag(111) substrate, which was first observed in Ref.~\cite{JAK1982} and analyzed from the band structure point of view in Ref.~\cite{Krasovskii99}. Because the initial states are strongly localized in the surface layer, the strong decay of the final state due to the gap in itself can hardly affect the intensity. Nevertheless, at energies of the gap the propagation of the outgoing photoelectron proceeds differently, which is illustrated by the energy-depth distribution of the probability density $\rho (z)$ in the final state $\ket{\Phi}$ in Fig.~\ref{fig2}(e). One can see a scattering resonance near 22 eV that is also visible in the line profile at 20.5 eV in Fig.~\ref{fig2}(f). The rapid modification of the final state in approaching the resonance gives rise to the observed intensity modulations. Notice, in particular, the nearly symmetric behavior of $\rho (z)$ around the Bi atom at around 20.5 eV where the intensity $I_z$ drops to zero.   

We will now show how the variations of $T_z$ and $T_x$ affect the photoelectron spin polarization. Along the $k_y$ direction the wave function $\Psi_{\rm +}$ [Fig.~\ref{fig1}] reads $\Psi_{\rm +} = g\xu + u\xd$, with $g(\rv)$ and $u(\rv)$ being even and odd upon reflection $x\to-x$, respectively, and $\chi^{\sigma}$ quantized along $x$. In this geometry, the intensity $I_z$ corresponds to spin-$\uparrow$ and the intensity $I_x$ to spin-$\downarrow$ photoelectrons, because $\mathcal{E}_z$ couples to $g(\rv)$ and $\mathcal{E}_x$ to $u(\rv)$. Hence, we expect the photoelectron spin polarization $P_x$ to have opposite sign for $I_z>I_x$ and $I_z<I_x$.

This is confirmed by the spin-resolved data in Fig.~\ref{fig3}(a). The measured polarization flips sign twice as a function of $\hbar\omega$ [Fig.~\ref{fig3}(e)], confirming that $I_z - I_x$ changes sign as expected from the spin-integrated data. Additional insight is gained from data acquired with $s$-polarized light in Fig.~\ref{fig3}(a)-(b). Along $k_y$, $s$-polarized light only couples to the even part $g(\rv)$ of $\Psi_{\rm +}$, as does the $z$-component of $p$-polarized light. Thus, if $I_z - I_x >0$ the sign of the photoelectron spin should coincide for $s$- and for $p$-polarized light, whereas opposite signs are expected if $I_z - I_x <0$. We find the former at $\hbar\omega =$~20~eV and the latter at $\hbar\omega =$~24~eV [Fig.~\ref{fig3}(a)-(b)], which implies $I_x > I_z$ between approximately $\hbar\omega =$~22~eV and 26~eV. 

Our one-step photoemission calculations of the photoelectron spin polarization in Figure~\ref{fig3}(c) confirm the sign change. We find that the spin polarization depends on the angle of light incidence $\theta$, because the relative magnitudes of $\mathcal{E}_x$ and $\mathcal{E}_z$ vary with $\theta$ [Fig.~\ref{fig3}(d)]. The spin polarization changes sign for $\theta =$~45$^\circ$ but for $\theta =$~73$^\circ$ it merely dips to 0. In the experiment a sign change is observed at $\theta =$~73$^\circ$. This discrepancy likely arises from uncertainties concerning the actual exciting field at the surface: the dielectric response is rather large at around $\hbar\omega =$~21~eV, so that the electric field at the surface may differ both from the field in the vacuum and from the field in the depth of the crystal \cite{Krasovskii:10}. However, in the studied $\hbar\omega$-range, the real part of the dielectric function $\epsilon_1$ of Ag is positive and shows a smooth behavior \cite{Hagemann:75}, so that we expect no qualitative influence of the dielectric response on the observed photon-energy dependence.   

In conclusion, we demonstrated that the photoelectron spin polarization of 
a spin-orbit-split surface state can strongly modulate and fully reverse upon 
only small changes in the excitation energy ($\delta \omega / \omega <$~10\%). 
Such a strong effect of the variation of the final state with energy will be of 
importance for spin-ARPES experiments on various spin-orbit coupled materials, 
such as topological insulators, Weyl semimetals, and Rashba systems. In contrast to previous photoemission studies, 
which focused {\it either} on the spin polarization {\it or} on the intensity (e.g., circular dichroism),
here the relation between the two observables, originating from the dipole selection rules, unambiguously points to the origin of the effect. Our findings, thereby, challenge the claim of a negligible role of the dipole operator 
stated in a previous work on the same material \cite{Chiang:12}.

Remarkably, our theory reproduces the nontrivial photon-energy-dependence of both the 
intensity and the spin polarization of the photocurrent. It also explains their origin 
as due to modulations of the final-state wave function at a Bragg gap in the unoccupied 
band structure. Details of the final state wave function are seen to have a major effect on the 
spin photocurrent even when spin-dependent scattering of the outgoing photoelectron does 
not lead to a spin polarization, that is when spin-orbit coupling in the final state is 
neglected. The latter effect -- spin rotation in the course of propagation \cite{Oepen1986} -- is 
known, both experimentally and theoretically, to be appreciable at kinetic energies of 
a few eV, but at small angles it rapidly vanishes at higher energies \cite{Samarin2007}. 
By contrast, the modulated photoelectron spin polarization observed in this work reflects the {\it intrinsic} spin properties of the probed state, namely its spatially
varying spin density, which is sampled differently depending on the shape 
of the final-state wave function. This is in contrast to the seemingly similar 
photon-energy-dependence of the circular dichroism in related systems \cite{Scholz:13}.

It is common belief that dichroic photoemission is an indispensable 
source of information about the spinor wave function. The vast majority of 
previous works considered circular dichroism, and a variety of 
interpretations have been put forward that related it to
the spin~\cite{Wang:11} or to a local orbital angular momentum~\cite{OAM-Park2012} 
in the initial state or to a final-state effect~\cite{Scholz:13}. No consensus appears to be reached as to how
the circular dichroism reflects the properties of the relativistic wave function. 
By contrast, the linear dichroism we report has a clear interpretation and originates 
from the symmetry properties of the spin-orbit coupling in the initial state in 
combination with the asymmetry of the experimental geometry. Hence, our study 
reinforces linear dichroism --as opposed to circular dichroism-- as a natural and efficient probe 
of spin-orbit coupling in the band structures of solids and surfaces \cite{Henk:04,Kuch:01}.

\section{Acknowledgments}
We gratefully acknowledge collaboration with Peter Kr\"uger, and thank Christian Ast, Simon Moser, and Hugo Dil for helpful discussions. This work was supported by the DFG through SFB1170 'Tocotronics' (Project A01) and by the Spanish Ministry of Economy and Competitiveness MINECO (Project No. FIS2016-76617-P).

\end{document}